\documentclass{pasj00}

\begin{document}
\SetRunningHead{Popovi\'c et al.}{Probing the physical properties of
the NGC 5548 BLR using Balmer lines}
\Received{2007/07/15}
\Accepted{2008/01/01}

\title{Probing the physical properties of the NGC 5548 Broad Line Region using Balmer lines}



%
 \author{%
   Luka \v C. \textsc{Popovi\'c},\altaffilmark{1}
   Alla I. \textsc{Shapovalova},\altaffilmark{2}
   Vahram H. \textsc{Chavushyan},\altaffilmark{3}
   Dragana \textsc{Ili\'c},\altaffilmark{4}
   Alexandr N. \textsc{Burenkov},\altaffilmark{2}
   Abelardo \textsc{Mercado}\altaffilmark{5}
   and
   Nikolay G.\textsc{Bochkarev}\altaffilmark{6}}
 \altaffiltext{1}{Astronomical Observatory, Volgina 7, 11160 Belgrade, Serbia}
 \email{lpopovic@aob.bg.ac.yu}

 \altaffiltext{2}{Special Astrophysical Observatory of the Russian
AS, Nizhnij Arkhyz, Karachaevo-Cherkesia 369167, Russia}
 \email{ashap@sao.ru}
 \email{ban@sao.ru}

 \altaffiltext{3}{Instituto Nacional de Astrof\'{\i}sica, Optica y
Electr\'onica, Apartado Postal 51, CP 72000, Puebla, Pue. M\'exico}
 \email{vahram@inaoep.mx}

 \altaffiltext{4}{Department of Astronomy, Faculty of Mathematics, University of
Belgrade, Studentski trg 16, 11000 Belgrade, Serbia}
 \email{dilic@matf.bg.ac.yu}

 \altaffiltext{5}{Facultad de Ingenier\'{\i}a, Universidad Nacional Aut\'onoma de Baja
California, 21280, Mexicali, B.C., M\'exico}
 \email{abel@astrosen.unam.mx}

 \altaffiltext{6}{Sternberg Astronomical Institute, Moscow, Russia}
 \email{boch@sai.msu.ru}


\maketitle

\begin{abstract}
We investigate the physical characteristics of the Broad Line Region
(BLR) of NGC 5548 using the Boltzmann plot (BP) method, that we
applied on the Balmer lines observed from 1996 to 2004. We find that
variability earlier detected in the lines, is also present in the
temperature parameter $A$, obtained from the BP method. Using the BP
temperature parameter $A$ to calculate temperature, we find that the
average temperature for the considered period was T$\approx$ 10000
K, and that varies from 5000 K (in 2002) to 15000 K (in 1998). This
variation correlates with the AGN-component of the optical continuum
flux ($r=0.85$) and may indicate the existence of an accretion disc in the BLR of
NGC 5548.
\end{abstract}

\section{Introduction}
The physics and kinematics in the Broad Line Region (BLR) of
Active Galactic Nuclei (AGN) is more complicated than in its
Narrow Line Region (NLR) or in gaseous nebulae (Osterbrock 1989;
Krolik 1999; Sulentic et al. 2000 and references  therein). In
contrast to the NLR where forbidden lines (e.g. [OIII] and [NII]
lines) can be used as emitting plasma diagnostics,  the physical
conditions in the BLR cannot be obtained using simple relations
between the line ratios. The pure recombination conditions cannot
be applied in BLRs, e.g. the flux line ratios are different than
expected in the case of recombination (e.g. in some AGN
$Ly\alpha$/H$\beta \approx$ 10, Osterbrock 1989).

 Several effects can result in such line flux ratios of
hydrogen lines, particularly the collisional excitation and
extinction effects. For over more than 30 years now, numerous papers
have been dedicated to this problem, see e.g. Netzer (1975), Ferland
\& Netzer (1979), Ferland et al. 1979, Kwan (1984), Rees et al.
(1989), Ferland et al. (1992), Shields \& Ferland (1993), Dumont et
al. (1998), etc. Dust is present in the host galaxy of an AGN (see
e.g. Crenshaw et al. 2002, Crenshaw et al. 2004, Gabel et al. 2005,
etc.), but it seems that in some cases it cannot explain the
measured line flux ratios. The classical studies point toward
photoionization, as the main heating source of the BLR emitting gas,
that may explain observed spectra of AGN (see e.g. Kwan \& Krolik
1981,  Osterbrock 1989, Baldwin et al. 1995, Marziani et al. 1996,
Baldwin et al. 1996, Ferland et al. 1998, Krolik 1999, Korista \&
Goad 2004). On the other hand, some authors e.g. Dumont et al.
(1998) have given favor to a non radiatively heated region that
contributes to the BLR line spectrum, which may be the case in the
BLR of NGC 5548. Therefore, the photoionization, recombination and
collisions could be considered as relevant processes that occur in
BLRs. At larger ionization parameters, recombination is more
important, but at higher temperatures or in the case of low
ionization parameters the collisional excitation becomes important
also (Osterbrock 1989). These two effects, as well as
radiative-transfer effects in Balmer lines, should be taken into
account when explaining the ratios of Hydrogen lines. Moreover, the
geometry and possible stratification in the BLR may affect the
continuum and line spectra of AGN (Goad et al. 1993).

Recently, Popovi\'c (2003, 2006ab) showed that in the BLR of some
AGN, the Balmer line ratios follow the Boltzmann plot (BP), which
indicates that the  population of the levels with $n\ge3$ follow
the Saha-Boltzmann equation. If the population of  excited levels
of the Balmer series can be described by Saha-Boltzmann equation,
one can determine the temperature of the region where these lines
are formed. This method, and also recently given method by Laor
(2006) allow us to ascertain information about the physical
(thermodynamical) properties of the BLR in a direct way  by
only measuring the broad line parameters (Ili\'c et al. 2007).

In order to test BP method as a tool for the temperature
diagnostic in the BLR, we investigate here the spectra of NGC
5548 observed from 1996 to 2004 (Shapovalova et al. 2004). First,
we found that BP method can be applied to the broad Balmer lines
of NGC 5548, and after that we determined the temperature
parameter $A$ using the BP method.

NGC 5548 is  one of the most intensively monitored Sy 1
galaxies (Shapovalova et al. 2004 and references therein) and the
physics of its BLR have been studied by many authors (e.g. Krolik
et al. 1991, Shields \& Ferland 1993, O'Brien et al. 1994, Dumont
et al. 1998, Vestergaard \& Peterson 2005, Korista \& Goad 2000,
2004). In particular, Dumont et al. (1998) discussed the physical
model of the NGC 5548 BLR, and found inconsistencies between the
photoionization model and the measured line fluxes, line flux
variations and the underlying continuum. The aim of this work is
to investigate the variations of the physical properties of the
NGC 5548 BLR using the BP method and Balmer lines, during a
period of 8 years.

\section{The BP method}

For plasma of the length $\ell$ that emits along the line of sight,
the flux (or the spectrally integrated emission-line intensity)
$F_{ul}$ of the transition from upper to lower level ($u\to l$) can
be calculated as (see also Popovi\'c 2006b):

\begin{equation}
F_{ul}={hc\over\lambda}g_{u}A_{ul}\int_0^\ell N_u(x)dx
\end{equation}
where $\lambda$ is transition wavelength, $g_u$  statistical
weight of the upper level, $A_{ul}$  transition probability,
$N_u$ is the number  density of emitters excited in upper
level which effectively contribute to the line flux (which are not
absorbed) and $h$ and $c$ are the well known constants (Planck
and speed of light). In general, the $N_u$ can be inhomogeneous
across the line of sight and also, the radiative self-absorption
can be present. But, assuming that population in the observed
region (in all layers) follows the Saha-Botzmann distribution one
can write

\begin{equation}
N_u(x)\approx {N_0 (x)\over Z}\exp(-E_u/kT(x)),
\end{equation}
where $Z$ is  the partition function, $N_0$ is  the total number
density  of radiating species, $E_{u}$ is the energy of the upper
level, $T$ is temperature and $k$ is the Boltzmann constant.

In the case of optically thin plasma\footnote{ Note here that
we cannot expect that plasma in the BLR is optical thin, further
in the text we will consider possible distribution of emitters
within the BLR, see Eq. (5)} with relatively small changes in the
density and temperature one can write (see, e.g., Griem 1997;
Konjevi\'c 1999)
\begin{equation}
F_{ul}={hc\over\lambda}g_{u}A_{ul}\int_0^\ell N_udx\approx
{hc\over\lambda}A_{ul}g_u\ell{N_0\over Z}\exp(-E_u/kT).
\end{equation}
For one line  series (as e.g. Balmer line series) if the population
of the upper energy  states ($n\ge3$)\footnote{ Note here that since
the emission deexcitation goes as $u\to l$ it is not necessary that
level $l$ has a Saha-Boltzmann distribution.}  adheres to a
Saha-Boltzmann distribution than one can determine their temperature
($T$), from a Boltzmann plot as
\begin{equation}
\log(F_n)=\log{F_{ul}\cdot \lambda\over{g_uA_{ul}}}=B-A{E_u},
\end{equation}
where $F_n$ is normalized flux, $B$ and $A$ are BP parameters.
Parameter $A$, defined as $A=\log_{10}(e)/kT$ (where $e\approx
2.718$), is the temperature indicator out of which we can estimate
the temperature of the region where these lines are formed.

On the other hand, one cannot expect the homogenous distribution of
the physical parameters and density of emitters across the line of
sight in an extensive BLR. But, if we still have that population
follows the Saha-Boltzmann equation, Eq. (1) can be written as:

\begin{equation}
F_{ul}={hc\over\lambda}g_{u}A_{ul}\int_0^\ell {N_0 (x)\over
Z}\exp(-E_u/kT(x)) dx
\end{equation}
and if we divide the BLR in  small layers with the same physical
conditions and emitter density, then we can write:

\begin{equation}
F_{ul}={hc\over\lambda}g_{u}A_{ul}\sum_{n=1}^n {N_0 (i)\over
Z}\exp(-E_u/kT(i)) \ell_i.
\end{equation}
If we assume that the temperatures across the BLR vary as
$T(i)=T_{av}\pm \Delta T_i$ and emitter density as
$N_0(i)=N_0^{av}\pm\Delta N_0(i)$, the Eq. (6) can be written as:

\begin{equation}
I_{ul}={hc\over\lambda}g_{u}A_{ul}{N_0^{av}\over Z}
\sum_1^n {{(1+\delta N_0 (i))}}\exp\left[-{E_u\over kT_{av}(1+ \delta T_i)}\right]\ell_i,
\end{equation}
where $\delta T_i=\Delta T_i/T_{av}$ and
 $\delta N_0=\Delta N_0/N^{av}_0$. If in a BLR the temperature and emitter
density does not vary too much,
i.e. $\Delta N_0/N_0<<1$ and $\Delta T_i/T_{av}<<1$, then the Eq. (7) becomes

\begin{equation}
I_{ul}={hc\over\lambda}g_{u}A_{ul}{N_0^{av}\over Z}
\exp(-E_u/kT_{av}) \ell .
\end{equation}
meaning that the Eq. (4) can be used to determine
$T_{av}$ in the BLR (see Popovi\'c 2006b).

Radiative transfer issues (e.g. high optical depths of the Hydrogen
lines) complicates the picture, but some additional tests are
possible  (see \S 5.2 and also Ili\'c 2007). Furthermore, the "Case
B" recombination line ratios can produce similar Boltzmann plots of
the Balmer line series in some AGN whose BLR is close to "Case B"
recombination (see Osterbrock 1989). Due to the physical conditions
(densities and temperatures)  in such BLRs the constant $A$ is too
small ($A<0.2$) and the Boltzmann plot cannot be applied for
diagnostics of the temperature even if the BP can be properly
applied (Popovi\'c 2003, 2006ab, Ili\'c et al. 2006).

We should emphasize that the BP method (Popovi\'c 2003, 2006ab) does
not take into account any {\it a priori} physics in the BLR, besides
that the Balmer lines are originating in the same  emitting region.
The method includes the intensities and transition parameters of
five Balmer lines, not only the ratio of two or three lines, as
usually considered. We would like to point out  that the BP is not
the Balmer decrement, i.e. the ratio of the Balmer lines. Also, with
this method, we concentrate on the thermodynamical state of the BLR
trying to see if the BP method works at all,  without assumptions on
any  "macroscopic" model caused by the "microscopic" physics.

\section{Observations and selection of the spectra}

Optical spectra of NGC 5548 were taken with the 6 m and 1 m
telescopes of SAO (Russia, 1996-2004) and at INAOE's 2.1 m telescope
at the Guillermo Haro Observatory (GHO) at Cananea, Sonora, Mexico
(1998-2004). They were obtained with a long slit spectrograph
equipped with CCDs. The typical wavelength range covered was from
4000 \AA\ to 7500 \AA , the spectral resolution was 4.5-15 \AA , and
the S/N ratio was $>$50 in the continuum near the H$\alpha$ and
H$\beta$. Spectrophotometric standard stars were observed every
night. Details concerning the spectroscopic observations are given
in Shapovalova et al. (2004).

Note that the spectra of NGC 5548 were obtained under the AGN
spectral monitoring program with the aim to study the variability
of the broad Balmer emission lines H$\alpha$ and H$\beta$. The
errors of the continuum at 5100\AA\  and line fluxes were about
3-5\%. From the spectral database collected during the monitoring
program of NGC 5548 ($\approx$150 spectra) we selected only 24
spectra (Table~\ref{tab1}), which satisfy the following:
\begin{enumerate}
\item good photometric conditions;
\item S/N $>$ 10 in the continuum near the H$\epsilon$ line;
\item spectrum covers the wavelength range from $\lambda4000\AA$ to
$\lambda7000\AA$;
\item the broad component of the Balmer lines from
H$\alpha$ to H$\epsilon$ are presented.
\end{enumerate}

The basic data of these observations are presented in Table~\ref{tab1}.

The spectrophotometric data reduction was carried out either with
the software developed at the SAO RAS by Vlasyuk (1993) and with
IRAF for the spectra obtained in Mexico. The image reduction process
included bias subtraction, flat-field corrections, cosmic ray
removal, 2D wavelength linearization, sky spectrum subtraction,
stacking of the spectra for every set-up, and flux calibration based
on standard star observations. We also remove the continuum of the
host galaxy.

The procedure of absolute calibration of the spectra is described in
details in Shapovalova et al. (2004) and will not be repeated here.

\section{Measurements of the broad line flux}

\begin{figure}
  \begin{center}
    \FigureFile(75mm,75mm){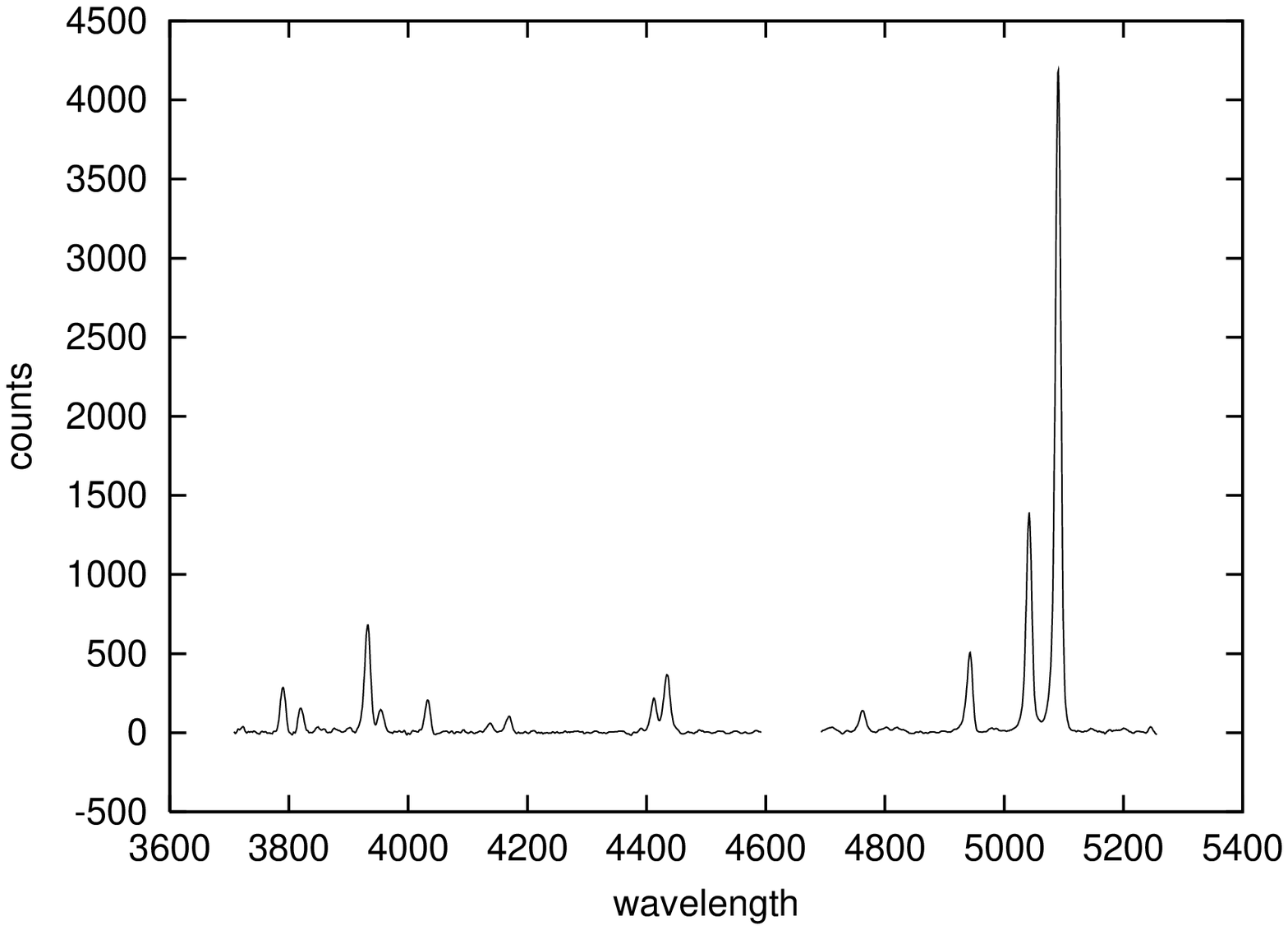}
    \FigureFile(75mm,75mm){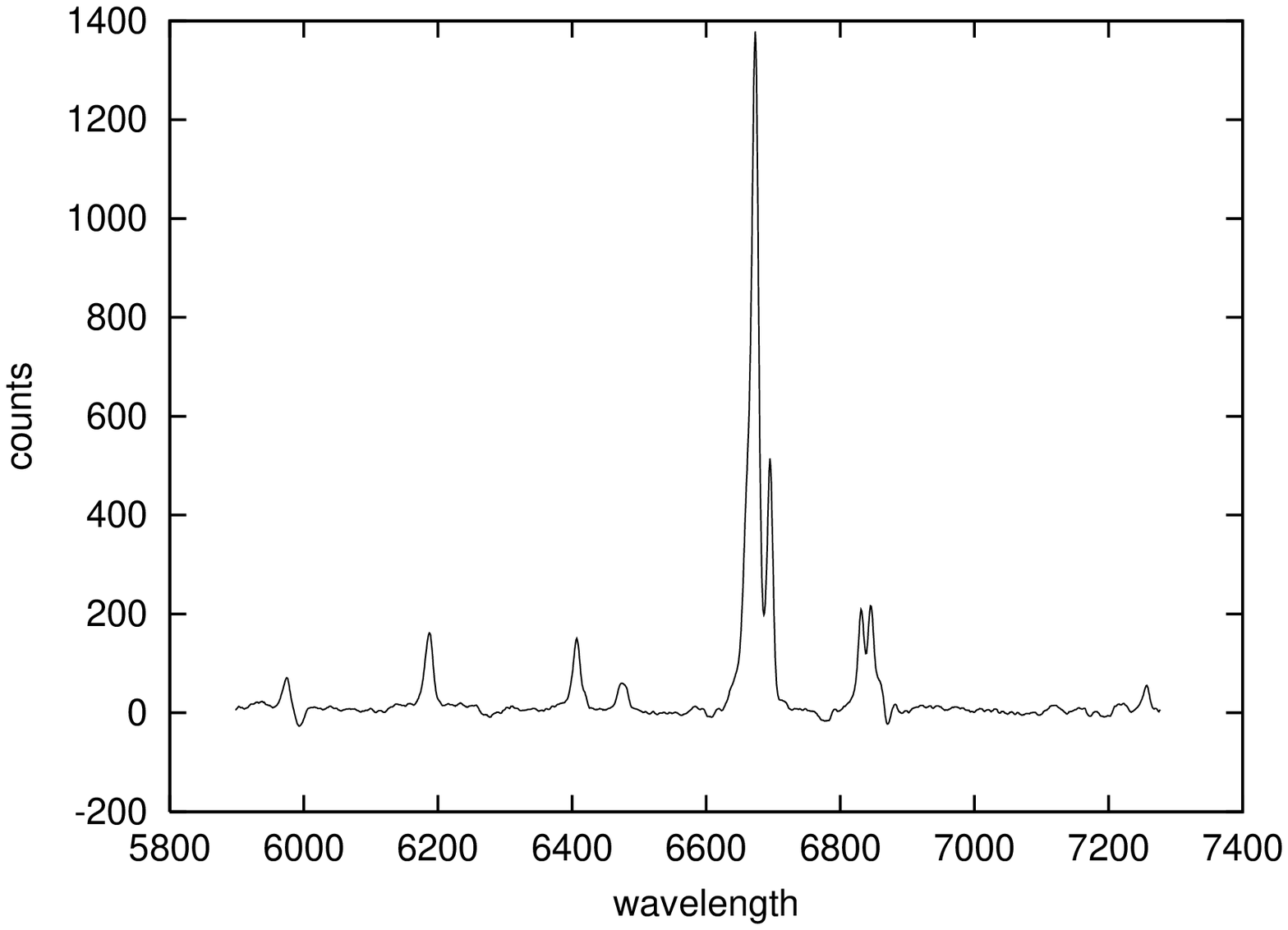}
  \end{center}
  \caption{The adopted narrow emission line template.} \label{fig1}
\end{figure}

One can expect that the BP can be applied only in the case of the
broad Balmer lines (Popovi\'c 2003), and first step was to subtract
the continuum and the narrow and satellite lines from the Balmer
lines. To estimate the contribution of these lines we used a
template of narrow and satellite lines (Figure~\ref{fig1}) estimated
from two independent measurements in the case of the minimum
intensity of the broad Balmer lines. In this template the Fe II
lines in the H$\beta$ spectral region are included. The optical Fe
II lines showed variability with smaller amplitude than that of the
H$\beta$ (around 50\%-75\%  of H$\beta$, Vestegaard \& Peterson
2005). Testing the contribution of the Fe II residuals (after
subtraction of the template), we concluded that it may contribute
about 2-5\% to the total measured line flux of the H$\beta$ line.

\begin{figure*}[ht]
\begin{center}
\includegraphics[width=50mm]{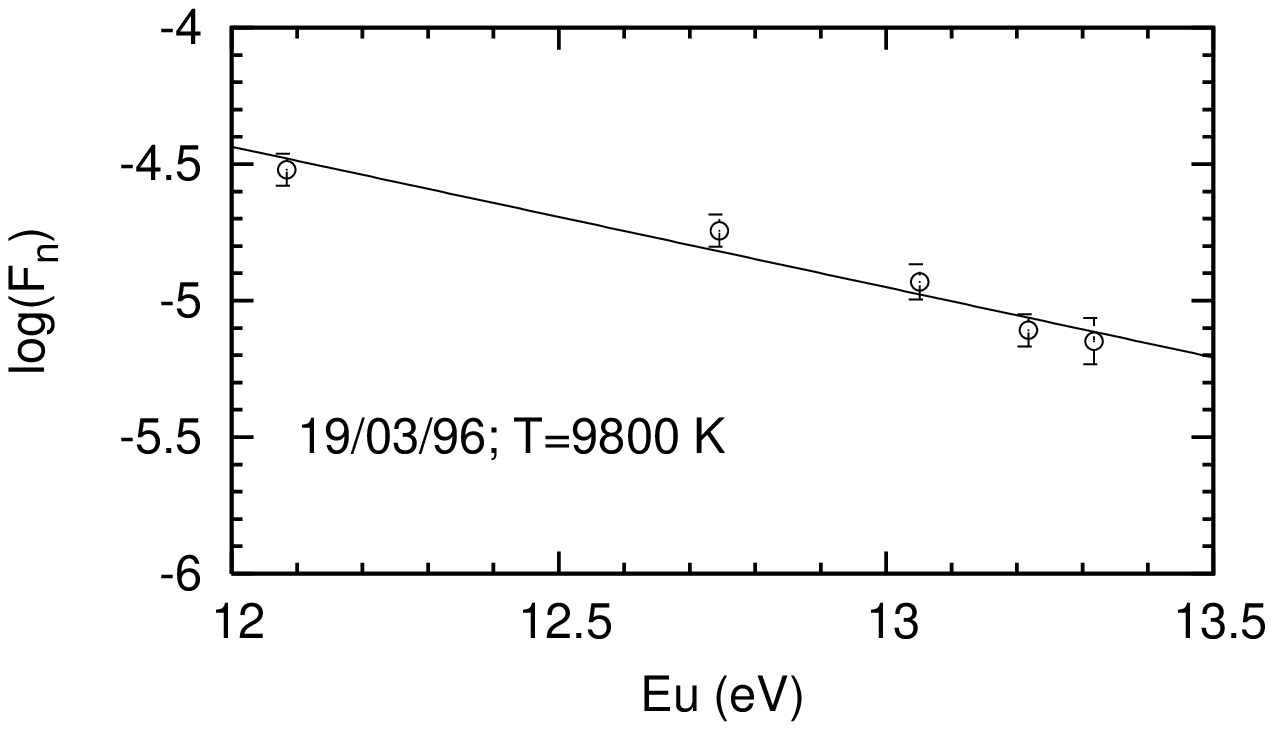}
\includegraphics[width=50mm]{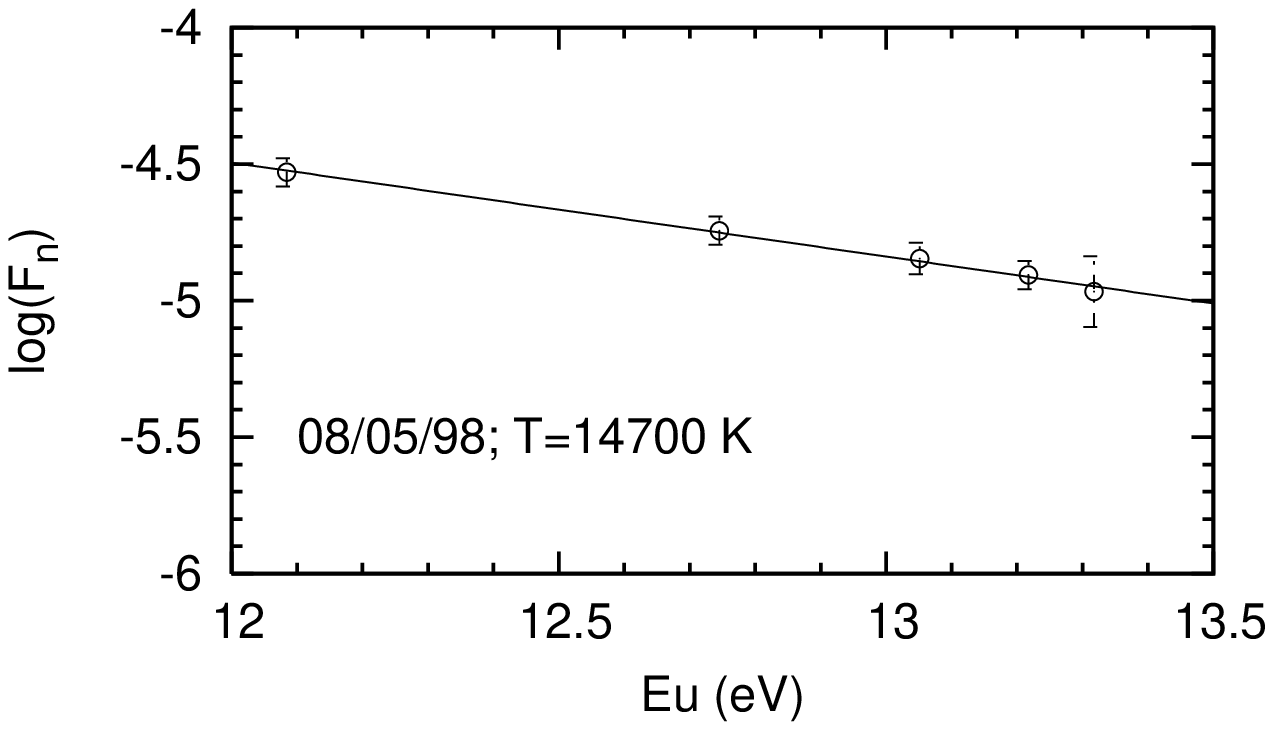}
\includegraphics[width=50mm]{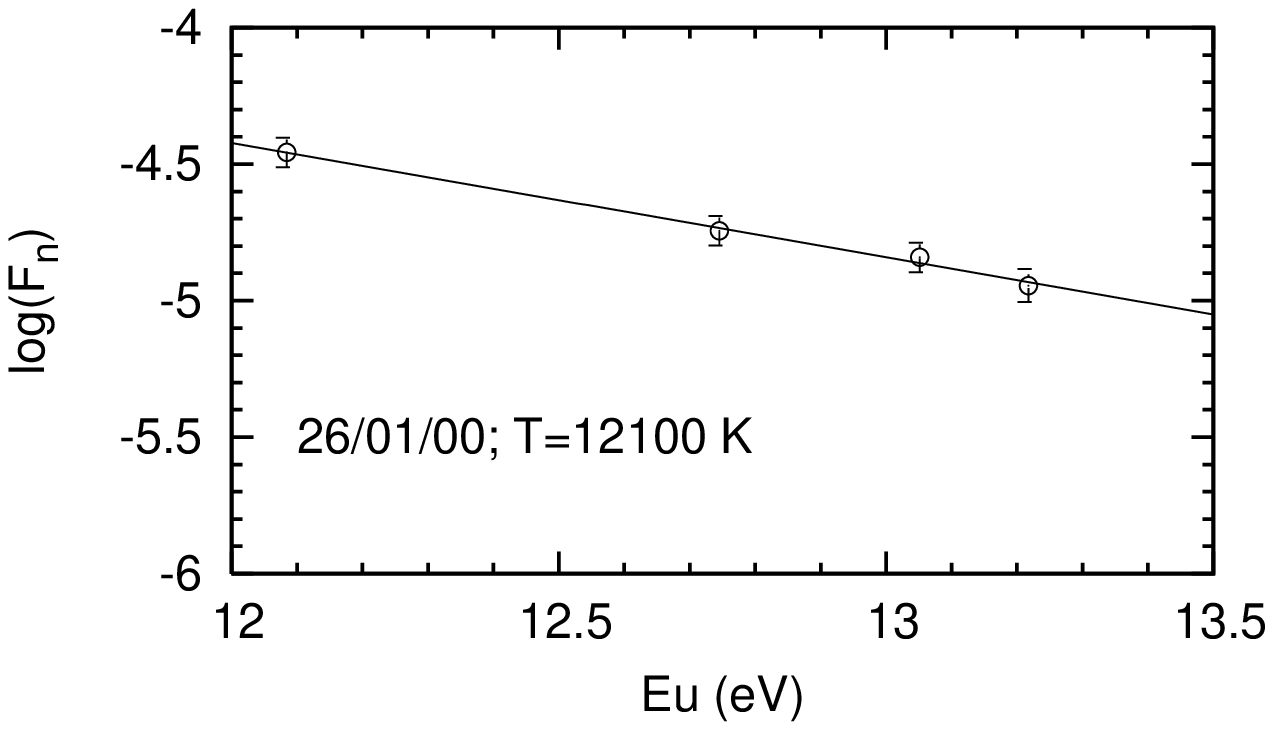}
\includegraphics[width=50mm]{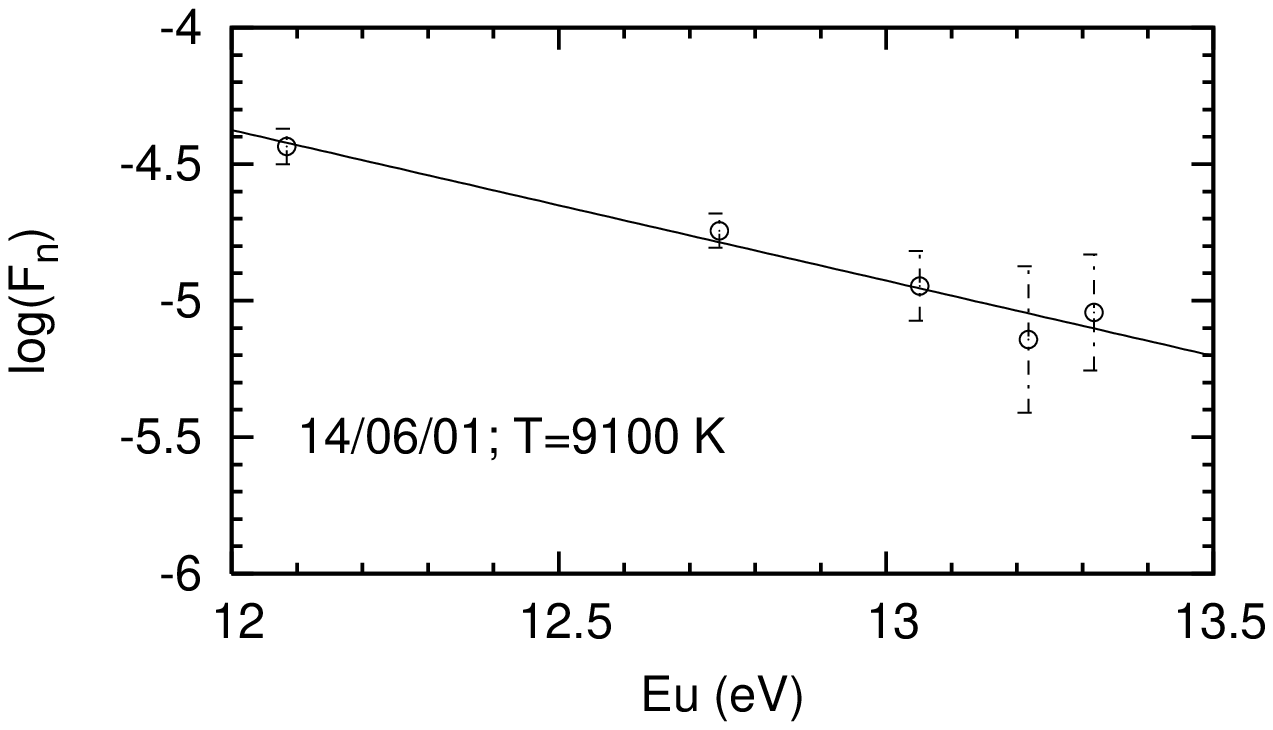}
\includegraphics[width=50mm]{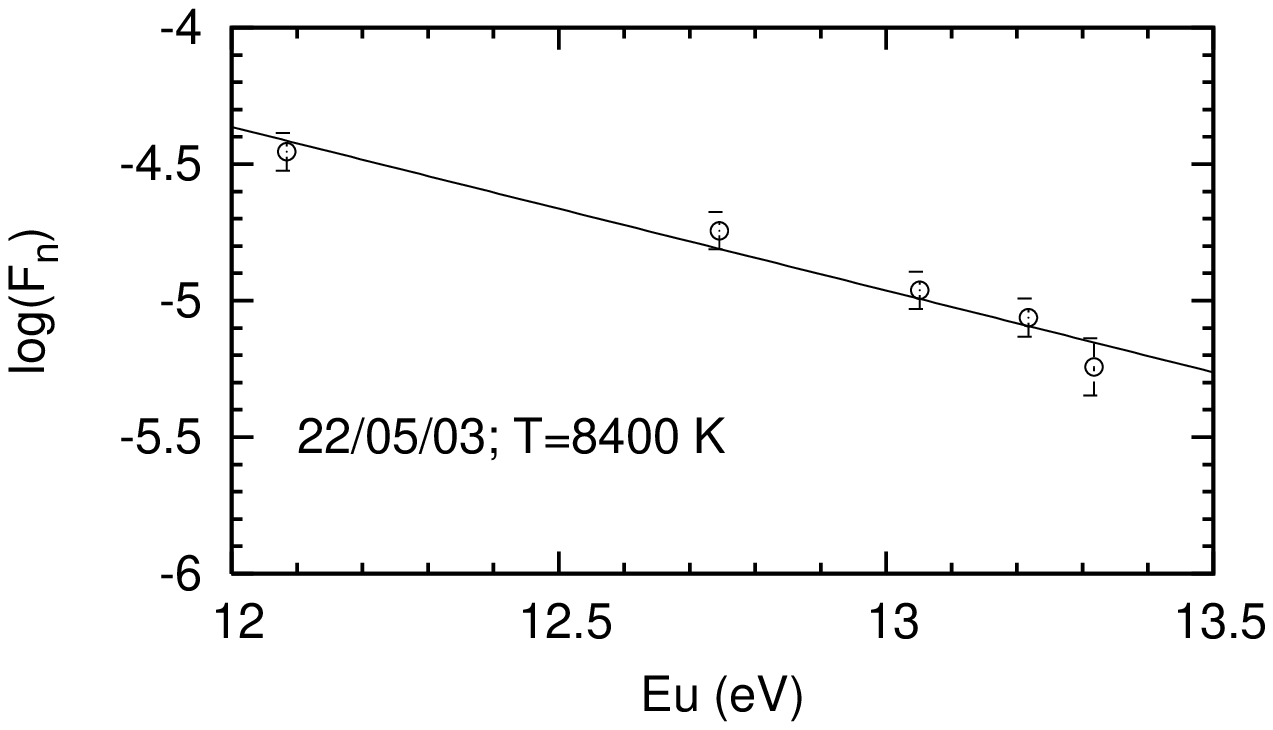}
\includegraphics[width=50mm]{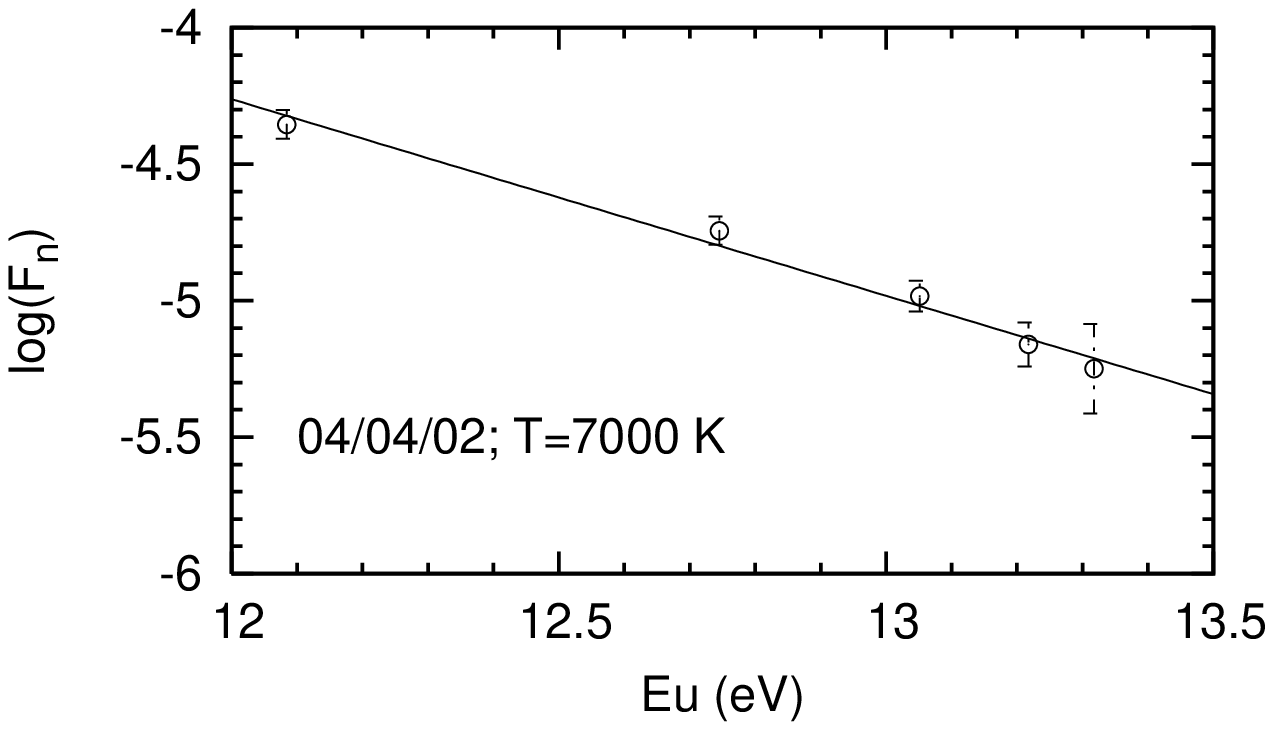}
\end{center}
\caption{The Boltzmann Plots of the Balmer line series of NGC 5548
from different periods.  $F_n$ is calculated using the Eq. (4) after
the normalization of all Balmer lines to the $H\beta$ flux ($F_{ul}
(H\beta)=1$).} \label{fig2}
\end{figure*}

Furthermore, we {\it a priori} take into account that the continuum
subtraction contributes to the error-bars within 10\% of the
measured fluxes. In general, we can expect this value in weak Balmer
lines (e.g. H$\varepsilon$), but in the case of the H$\alpha$ and
H$\beta$ lines, it should be even smaller.

The fluxes of the Balmer lines were measured several times, but it
is important that independent measurements were performed by two
persons. Then, the error-bars were calculated as:

\begin{equation}
\Delta F_{i}=\Delta F_{i}^{mes}+0.1\cdot F_{i}^{mes}
\label{eqn:flux}
\end{equation}
where $F^{mes}_i$ is the measured flux, $\Delta F_{i}^{mes}$ is
statistical error (within $1\sigma$) obtained from several
measurements and 0.1$F_{i}^{mes}$ is taken to be the error of the
continuum and narrow line template subtraction. The flux ratios of
the Balmer lines and the fluxes of the H$\beta$ from different
periods are given in Table~\ref{tab2}. The error-bars presented in
the Table~\ref{tab2} were obtained as (see e.g. Agekyan 1972,
 Bevington \& Robinson 2003):

\begin{equation}
\Delta R_{i}=R_i\cdot\sqrt{({\Delta
F_{i}\over{F_{i}^{mes}}})^2+({\Delta
F_{H\beta}\over{F_{H\beta}^{mes}}})^2}\label{eqn:flux_ratio}
\end{equation}
where $R_i$ is the ratio of $F_i/F_{H\beta}$.

In Figure~\ref{fig2} the error-bars  ($\Delta log_{10}(F_n)$) have
been calculated as:

\begin{equation}
\Delta log_{10}(F_n)= {\Delta F_i\over{F_i^{mes}\times \log_{e}10}}.
\label{eqn:error}
\end{equation}

One can conclude from Eq. (4) that the absolute scale is not
important for the BP analysis, which uses the slope to determine the
temperature coefficients ($A$),
 but one should re-scale fluxes of Balmer lines for the same factor
as well as the $\Delta F_i$. The parameters $A$ are obtained from
the best fit of the measured values and we give asymptotic standard
error $\Delta A$ (within 1$\sigma$) in Table~\ref{tab2}.

We should note here that we used only those observed spectra where
the spectral region from H$\varepsilon$ to H$\alpha$ was covered,
except for the two spectra (see Table~\ref{tab2}) that covered only
the interval from H$\delta$ to H$\alpha$. We used the aforementioned
spectra in order to have the data in years 2000 and 2001.

\begin{figure}
  \begin{center}
    \FigureFile(7.5cm,75mm){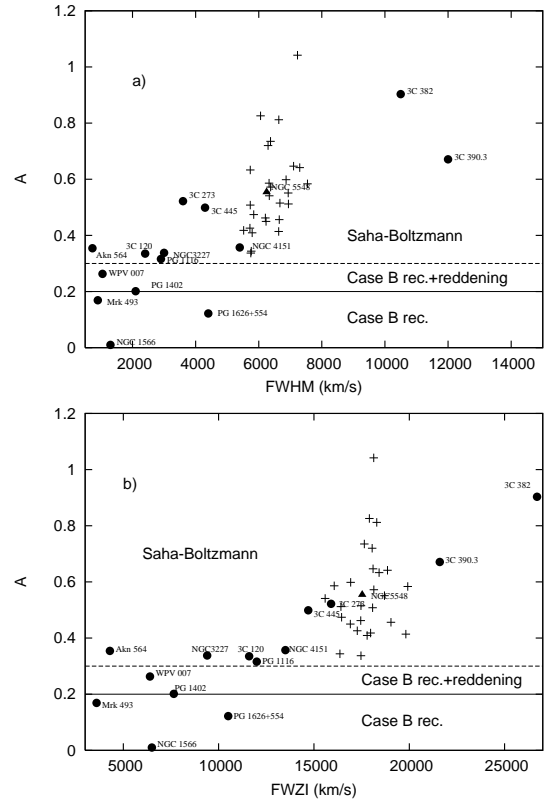}
  \end{center}
\caption{The graph $A$=f(FWHM,FWZI). The data from different periods
are denoted with crosses, while the averaged value for NGC 5548
is presented with full triangles. The data presented with full
circles are taken from Popovi\'c (2003). } \label{fig3}
\end{figure}

\begin{figure}
  \begin{center}
    \FigureFile(80mm,75mm){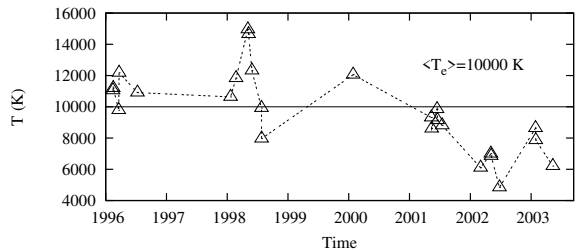}
  \end{center}
\caption{The temperature variation  from 1996 till 2004.}
\label{fig4}
\end{figure}

\begin{longtable}{cccccccc}
 \caption{The basic data of the spectroscopic observations
 of NGC 5548. The column "Code" refers to the telescope used for the observations:
 L - 6m telescope SAO RAS (Russia), GH - 2.1m telescope GHO
(Cananea, Mexico).} \label{tab1}
 \hline
 File     & UT-Date & JD         & Code & Aperture & Sp.range& Resolution& Seeing \\
 (spectra)&         & (2400000+) &       & (arcsec) & ({\AA}) & ({\AA})   & (arcsec)\\
\endfirsthead
\endhead
  \hline
\endfoot
  \hline
\endlastfoot
  \hline
960213 & 1996Feb14 & 50127.579 & L  & 1.5"x6.0" &  3100-7200  & 6 &   2 \\
960214 & 1996Feb14 & 50128.432 & L  & 1.5"x6.0" &  3100-7200  & 6 &   3 \\
960319 & 1996Mar19 & 50162.382 & L  & 2.0"x6.0" &  3600-7400  & 8 &   3 \\
960321 & 1996Mar21 & 50164.395 & L  & 2.0"x6.0" &  3600-7400  & 8 &   3 \\
960710 & 1996Jul10 & 50275.283 & L  & 2.0"x6.0" &  3600-7000  & 8 &   1.4\\
980120 & 1998Jan21 & 50834.632 & L  & 2.0"x6.0" &  3800-7600  & 8 &   3 \\
980222 & 1998Feb23 & 50867.535 & L  & 2.0"x6.0" &  3800-7600  & 8 &   2 \\
980504 & 1998May04 & 50938.293 & L  & 2.0"x6.0" &  3700-7600  & 8 &   2.5\\
980508 & 1998May08 & 50942.461 & L  & 2.0"x6.0" &  3700-7700  & 8  &  2 \\
980725 & 1998Jul26 & 51020.703 & GH & 2.5"x6.0" &  3960-7231  & 15 &   2.2\\
980726 & 1998Jul27 & 51021.712 & GH & 2.5"x6.0" &  3930-7219  & 15 &   2.5\\
200126 & 2000Jan27 & 51570.959 & GH & 2.5"x6.0" &  4070-7350  & 15 &   2.5\\
210513 & 2001May14 & 52043.859 & GH & 2.5"x6.0" &  3980-7300  & 15 &   1.8\\
210613 & 2001Jun14 & 52074.772 & GH & 2.5"x6.0" &  4022-7330  & 15 &  1.8\\
210614 & 2001Jun15 & 52075.738 & GH & 2.5"x6.0" &  4010-7330  & 15 &   1.8\\
21-07  & 2001Jul16 & 52107.320 & L  & 2.0"x6.0" &  3630-8050  &  8 &   2\\
220304 & 2002Mar05 & 52338.780 & GH & 2.5"x6.0" &  3976-7305  & 15 &   2\\
220404 & 2002Apr05 & 52369.850 & GH & 2.5"x6.0" &  3976-7305  & 15 &   2\\
220405 & 2002Apr06 & 52370.780 & GH & 2.5"x6.0" &  3976-7305  & 15 &   2\\
220624 & 2002Jun24 & 52450.420 & L  & 2.0"x6.0" &  3800-7600  &  8 &   2\\
230126 & 2003Jan27 & 52666.96  & GH & 2.5"x6.0" &  3976-7305  & 15 &   2.5\\
230127 & 2003Jan28 & 52667.930 & GH & 2.5"x6.0" &  3976-7305  & 15 &   2.5\\
230325 & 2003Mar26 & 52724.900 & GH & 2.5"x6.0" &  3976-7305  & 15 &   3.5\\
230522 & 2003May23 & 52782.750 & GH & 2.5"x6.0" &  3976-7305  & 15 &  3.5\\
\end{longtable}

\begin{longtable}{ccccccc}
 \caption{The measured flux of the Balmer lines  of NGC
5548, temperature parameter $A$.} \label{tab2}
 \hline
Spectra & F$_{H\alpha}$/F$_{H\beta}$ & F$_{H\gamma}$/F$_{H\beta}$ &
F$_{H\delta}$/F$_{H\beta}$ & F$_{H\varepsilon}$/F$_{H\beta}$ &
F$_{H\beta}\rm \ (erg\cdot cm^{-2}s^{-1})$ & A\\
\endfirsthead
  \hline
\endhead
  \hline
\endfoot
  \hline
\endlastfoot
  \hline
960213& 3.405$\pm$0.406& 0.396$\pm$0.047& 0.174$\pm$0.021& 0.071$\pm$0.027& (7.998$\pm$0.950)E-13&0.456$\pm$0.095\\
960214& 3.689$\pm$0.625& 0.456$\pm$0.074& 0.148$\pm$0.025& 0.091$\pm$0.016& (7.598$\pm$1.220)E-13&0.450$\pm$0.085\\
960319& 3.647$\pm$0.492& 0.341$\pm$0.051& 0.133$\pm$0.018& 0.077$\pm$0.015& (7.281$\pm$0.983)E-13&0.515$\pm$0.068\\
960321& 3.126$\pm$0.384& 0.386$\pm$0.048& 0.187$\pm$0.029& 0.072$\pm$0.016& (7.055$\pm$0.864)E-13&0.414$\pm$0.100\\
960710& 4.203$\pm$0.569& 0.465$\pm$0.063& 0.204$\pm$0.027& 0.084$\pm$0.011& (9.891$\pm$1.322)E-13&0.462$\pm$0.082\\
980120& 3.672$\pm$0.446& 0.393$\pm$0.048& 0.220$\pm$0.038& 0.062$\pm$0.008& (8.420$\pm$1.023)E-13&0.474$\pm$0.138\\
980222& 3.624$\pm$0.429& 0.376$\pm$0.072& 0.210$\pm$0.027& 0.082$\pm$0.023& (9.694$\pm$1.146)E-13&0.426$\pm$0.075\\
980504& 3.523$\pm$0.398& 0.401$\pm$0.048& 0.201$\pm$0.024& 0.125$\pm$0.042& (9.650$\pm$1.090)E-13&0.337$\pm$0.013\\
980508& 3.571$\pm$0.427& 0.416$\pm$0.055& 0.212$\pm$0.025& 0.117$\pm$0.035& (7.998$\pm$0.950)E-13&0.344$\pm$0.014\\
980725& 3.961$\pm$0.471& 0.417$\pm$0.049& 0.167$\pm$0.020& 0.072$\pm$0.009& (9.969$\pm$1.179)E-13&0.508$\pm$0.088\\
980726& 3.727$\pm$0.438& 0.338$\pm$0.047& 0.153$\pm$0.018& 0.039$\pm$0.012& (9.235$\pm$1.083)E-13&0.633$\pm$0.177\\
200126& 4.223$\pm$0.528& 0.420$\pm$0.052& 0.194$\pm$0.027&  -             & (7.167$\pm$0.886)E-13&0.418$\pm$0.022\\
210513& 5.718$\pm$1.134& 0.314$\pm$0.099& 0.109$\pm$0.049& 0.146$\pm$0.096& (2.900$\pm$0.437)E-13&0.586$\pm$0.161\\
210613& 4.493$\pm$0.622& 0.414$\pm$0.056& 0.150$\pm$0.077&  -             & (3.874$\pm$0.524)E-13&0.512$\pm$0.073\\
210614& 4.439$\pm$0.664& 0.330$\pm$0.097& 0.123$\pm$0.076& 0.098$\pm$0.048& (3.408$\pm$0.499)E-13&0.551$\pm$0.070\\
21-07 & 4.028$\pm$0.521& 0.390$\pm$0.065& 0.166$\pm$0.024& 0.055$\pm$0.009& (5.901$\pm$0.749)E-13&0.572$\pm$0.127\\
220304& 5.411$\pm$0.784& 0.289$\pm$0.042& 0.125$\pm$0.022& 0.035$\pm$0.012& (3.616$\pm$0.520)E-13&0.826$\pm$0.143\\
220404& 5.347$\pm$0.654& 0.303$\pm$0.039& 0.118$\pm$0.022& 0.061$\pm$0.023& (7.998$\pm$0.950)E-13&0.720$\pm$0.049\\
220405& 5.619$\pm$0.895& 0.329$\pm$0.052& 0.126$\pm$0.023& 0.057$\pm$0.019& (3.311$\pm$0.523)E-13&0.735$\pm$0.064\\
220624& 4.781$\pm$1.127& 0.057$\pm$0.017& 0.062$\pm$0.011& 0.029$\pm$0.012& (2.691$\pm$0.472)E-13&1.042$\pm$0.276\\
230126& 6.546$\pm$0.984& 0.292$\pm$0.066& 0.168$\pm$0.032& 0.140$\pm$0.048& (2.421$\pm$0.364)E-13&0.584$\pm$0.117\\
230127& 6.794$\pm$1.176& 0.208$\pm$0.040& 0.162$\pm$0.030& 0.130$\pm$0.032& (2.136$\pm$0.369)E-13&0.641$\pm$0.167\\
230325& 4.532$\pm$0.707& 0.238$\pm$0.039& 0.160$\pm$0.026& 0.079$\pm$0.031& (3.252$\pm$0.505)E-13&0.586$\pm$0.077\\
230522& 4.248$\pm$0.677& 0.318$\pm$0.050& 0.148$\pm$0.024& 0.062$\pm$0.015& (4.782$\pm$0.758)E-13&0.599$\pm$0.073\\

\hline

\end{longtable}

\subsection{Reddening}

The reddening effect can influence the Balmer line ratio (e.g.
Crenshaw \& Kraemer 2001; Crenshaw et al. 2001; 2002; Popovi\'c
2003) and consequently the temperature parameter obtained by the BP.
In the case of NGC 5548 the Galactic reddening is negligible E(B-V)
= 0.020 mag (Burstein \& Heiles 1982; Schlegel et al. 1998), and
here it is not considered. Concerning the intrinsic reddening, in
first approximation we can adopt that it might be 30\% - 40\%, as it
was given in Popovi\'c (2003). But, since we are investigating the
changes in the BP during an interval the intrinsic reddening can be
neglected as it should not vary too much in a relatively short
period of around 8 years.

\subsection{Velocity Measurements}

These measurements were performed in order to present  $A$ as a
function of Full Width Half Maximum (FWHM) and Full Width Zero
Intensity (FWZI). The graphs $A$ vs. FWHM and FWZI are useful
because of:

(i) the Balmer line ratios are velocity  dependent in AGN (Stirpe
1990,1991) and this may be related to the physical conditions
(temperature, density and kinematics) as well as to the
radiative transfer effects;

(ii) illustration of the physical conditions in the BLR plasma. As
it was pointed out by Popovi\'c (2003) if $A<0.3$, the recombination
"Case B" plus the intrinsic reddening may explain BP. The "Case B"
recombination of Balmer lines  can bring the $\log(F_n)$ {\it vs}
$E_u$ as a linear decreasing function (see Table
4.4 in Osterbrock 1989).

In order to compare the temperature parameter $A$ with the velocity
of gas, very high data quality is required. The data quality is
different for different spectra of NGC 5548. Consequently, for
velocity measurements we used only  the H$\alpha$ and H$\beta$
lines. We first normalized the broad profiles of these lines and
converted the wavelength axis to the velocity scale $X =(\lambda -
\lambda_0)/\lambda_0$  (the same as was given in Popovi\'c 2003). We
measured FWHM and FWZI of both lines, and we calculated the average
values for each  spectrum.

\section{Results and discussion}

\subsection{BP of NGC 5548}

The results of our investigations are presented in
Figures~\ref{fig2}-\ref{fig4}, and in Table~\ref{tab2}.

One can see in Figures~\ref{fig2} and ~\ref{fig3} that BPs graphics
indicate that the population of the excited levels from the Balmer
series follows the Saha-Boltzmann distribution. We found that in all
considered periods, the temperature parameter is $A>0.3$. It means
that the recombination "Case B" plus an intrinsic reddening cannot
explain the line flux ratio of the Balmer lines. Also, we
investigate the dependence of the temperature coefficient as a
function of FWHM and FWZI (Figure~\ref{fig3}). As one can see from
Figure~\ref{fig3}, there is no significant correlation between
widths and $A$. That is expected since macroscopic bulk motions are
mainly constrained by kinematics, probably an accretion disc as it
was suggested by Shapovalova et al. (2004).  On the other hand, it
may indicate that changes in the self-absorption in the Balmer lines
has no significant variation during the period. The averaged value
obtained from all periods (full triangle in Figure~\ref{fig3}) is in
good agreement with the trend of the AGN sample given in Popovi\'c
(2003). Such a trend (increasing $A$ with FWHM and FWZI) may
indicate that physical conditions are dependent on kinematical
characteristics of the BLR. This may be connected with distances
from the massive black hole, e.g. the broader Balmer lines indicate
higher macroscopic bulk motions which can be expected in the regions
closer to the massive black hole.

Also, we could estimate the temperature using Eq. (3) \& (4) from:

\begin{equation}
T\approx {5060\over{A}}\ [\rm{K}]. \label{eqn:temp}
\end{equation}

In Figure~\ref{fig4} we plotted the temperature variation in the
considered period: as one can see the averaged temperature is
$\approx$10000 K, and is changing by around 50\% in the considered
period. The maximum value of the temperature was reached in 1998 and
minimum in 2002. The obtained temperatures agree well with the ones
expected in the BLR.

On the other hand, from the BP fits, we obtained the parameter B. As
it can be seen from Eqs. (1)-(4) the parameter B depends on the
number density of radiating species and the partition function
($B=\log(hcN_0/Z)$). Both values are the same for the Balmer line
series at one moment, but they are changing during the time, since
we expect that the physical properties in the emitting BLR plasma of
NGC 5548 are changing (seen as variability in the broad spectral
lines). In principle the partition function depends on the
temperature and the level configuration (in our case Z is given for
$n=2$). It is hard to find some real information about variation in
the number density  in the BLR plasma using the B parameter, but
still one can have impression about magnitude of variations of these
two quantities. Taking into account Eqs. (1) - (4) and also the fact
that we normalized Balmer line fluxes to the H$\beta$ flux, we can
write
 $\log(N_0/Z)\sim B_{E_l}+\log(F_{H\beta})$ ($h$ and $c$ are
constants, and $B_l$ is intercept corresponding to the energy of the
lower level, since Z is given for $n=2$, $E_l=10.2$ eV). Using this
relation we found the difference between maximal and minimal value
of $\log(N_0/Z)$ to be
$\Delta\log(N_0/Z)=\log(N_0/Z)_{max}-\log(N_0/Z)_{min}\sim 0.9$.
This can be expected, e.g. if we roughly approximate that $Z$ stay
constant, then maximal variation of number density is $\sim$ 8
times, this is in an agreement with observed magnitude of variation
in spectral lines of NGC 5548 (e.g. the flux of the H$\beta$ was
changing $\sim$ 5 times.)

\subsection{Continuum vs. Temperature}

Here we start from the fact that the response of the broad lines to
continuum variations in NGC 5548 suggests that BLR is very close to
the continuum source (Ferland et al. 1992), or it may even indicate
that a part of optical continuum is originating in the BLR as well.
We assume  that the continuum is originating in an accretion disc,
e.g. Kong et al. (2004) explained the variation in the optical
spectral index with the variation in accretion parameters of the
inner part of the disc. On the other hand, the Balmer lines may also
be emitted from outer part of the accretion disc (Shapovalova et al.
2004), then the temperature variation measured from BP should be
connected with the continuum variation. Therefore, we investigate
the continuum intensity as a function of obtained temperatures. In
an accretion disc, temperature depends on the radius (see e.g.
Shakura \& Sunyaev 1973) but if  the continuum is emitted mostly
from inner parts of the disc and Balmer lines mostly from outer
parts, one can approximate that $T_{\rm c}$ correlates with the BLR
temperature, i.e. $T_{\rm c} \sim T_{\rm BLR}$.

To find the correlation, we measured the flux in a window from 4240
\AA \ to 4260 \AA\ (rest-frame wavelengths) and calculated an
average value for 4250 \AA\ as well as in the spectral range from
5090 \AA\ to 5110 \AA\ (rest wavelengths) with an average value at
5100 \AA .

\begin{figure}
  \begin{center}
    \FigureFile(75mm,75mm){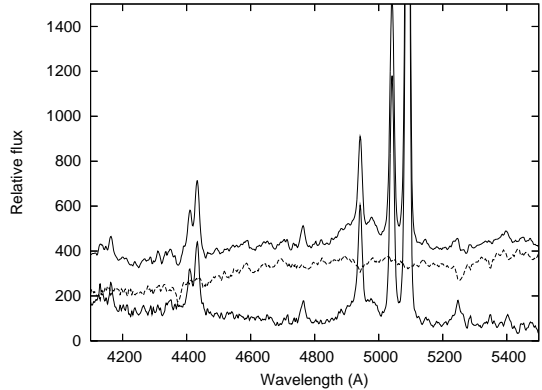}
  \end{center}
\caption{The continuum flux of NGC 5548 in minimum before (top solid line) and
after (bottom solid line) subtraction of the host galaxy continuum (dashed line).
 } \label{fig5}
\end{figure}

 In order to have only the AGN-component of the optical continuum,
one should subtract the continuum of the host galaxy from the observed
continuum of NGC 5548 nucleus. For this purpose, the contribution
of the host galaxy at wavelengths 4250\AA\ and 5100\AA\ was
obtained as following:

a) according to the paper by Romanishin et al. (1995, Fig.~5) the
host galaxy contribution in the AGN continuum of NGC 5548 at 5100
\AA\ in the aperture $5''\times7.5''$  is
 $F(5100)=(2.9\pm0.54)\cdot10^{-15} {\rm erg\cdot cm^{-2}\cdot sec^{-1}\cdot\AA^{-1}}$.
Since our spectra were taken with the smaller aperture
$2.5''\times 6''$, it is necessary  to determine the correction
for the aperture effect. We compared the continuum fluxes of NGC
5548 at 5100\AA\  using spectra obtained in the same nights with
our aperture $2.5''\times 6''$ and with the aperture
$5''\times7.5''$ of Peterson et al. (2002). The spectral
observations during 13 different nights were used. The average
correction for aperture effect is $\Delta F = F((5''\times7.5''
)-F(2.5''\times 6'' ) = 0.35\cdot10^{-15} {\rm erg\cdot
cm^{-2}\cdot sec^{-1}\cdot\AA^{-1}}$. The host galaxy
contribution at 5100\AA\ with aperture $2.5''\times 6''$ is
F(5100)gal = $(2.55\pm0.54)\cdot10^{-15} {\rm erg\cdot
cm^{-2}\cdot sec^{-1}\cdot\AA^{-1}}$;

b) For the determination  of the host galaxy  contribution to the
continuum flux of NGC 5548 at the wavelength 4250\AA\ the spectrum
of the normal E galaxy NGC 4339 was used. The spectral
observations of NGC 4339 were taken  with the 2.1 m telescope with
the same aperture $2.5''\times 6''$ and with the same spectral
resolution ($\approx8\AA$) as the spectrum of NGC 5548 in the
minimum activity state. After that we compared the spectra of NGC
4339 with the spectrum NGC 5548 in the minimum state (June 4,
2002). Such a comparison is justified because the central colors
of the host galaxy NGC 5548 (B-V$\approx$0.9; V-I$\approx$1.2),
are similar to the colors of an elliptical galaxy (Romanishin et
al. 1995). We varied the contribution from NGC 4339 at 5100\AA\
from 50\% to 100\% of that in the observed spectrum of the NGC
5548 and found, using a good substraction of
 the absorption
lines, that in the minimum activity of NGC 5548 the host galaxy
contributes around 70\% to the total continuum flux at 5100\AA\
(see Figure \ref{fig5}). From Shapovalova et al. (2004) the
observed flux at  5100\AA\ in NGC 5548 for the minimum spectrum
in aperture $2.5''\times 6''$ is F(5100)$\approx3.7\cdot10^{-15}
{\rm erg\cdot cm^{-2}\cdot sec^{-1}\cdot\AA^{-1}}$; and then the
host galaxy of NGC 5548 flux is
F(5100)$\approx(3.7\cdot0.7)\cdot10^{-15}=2.5\cdot10^{-15} {\rm
erg\cdot cm^{-2}\cdot sec^{-1}\cdot\AA^{-1}}$, that coincides with
the value obtained by us, using the data from Romanishin et al.
(1995).

We found that for NGC 4339 the fluxes ratio at 5100\AA\ and
4250\AA\ is F(5100)/F(4250)=1.45. Supposing that the continuum of
the host galaxy NGC 5548 is similar to that in NGC 4339 we found
that the flux of  host galaxy NGC 5548 at $\lambda$4250\AA\ is
F(4250)=F(5100)$/1.45=(1.76\pm0.54)\cdot10^{-15} {\rm erg\cdot
cm^{-2}\cdot sec^{-1}\cdot\AA^{-1}}$; Then we obtained the
AGN-component continuum flux of the NGC 5548 at $\lambda$4250\AA\
and 5100\AA\ by substraction of the continuum flux of the host
galaxy. The measured fluxes only of the AGN-component are given
in Table~\ref{tab1}.

Note here that Peterson et al. (1995) showed that in the case of NGC
5548 the NLR  has the same surface-brightness distribution as the
PSF (i.e the NLR is point-like source $\sim$2"), and that the
point-source correction factor is always equal to one. Consequently,
the broad-line/narrow-line flux ratio should be independent of the
aperture and the seeing, at last for (slit)$\ge$ 2", that is the
case in our sample of observations.

The total continuum flux ($F_t(\lambda)$) emitted at $\lambda$ can
be represented as a sum of the BLR ($F_{BLR}$), central source
($F_c(\lambda)$) and stellar continuum ($F_s(\lambda)$). As we noted
above, the stellar continuum is subtracted and then:

\begin{equation}
F_t(\lambda,T)=F_c(\lambda,T_c)+F_{BLR}(\lambda,T_{\rm BLR}).
\label{eqn:IIIa}
\end{equation}

Assuming that the optical continuum is originating far enough from
the centre of an accretion disc, one can use the next relation for
the central source of continuum (see Peterson 2004, Eq. 4.6 and
corresponding discussion)

\begin{equation}
F_c(\lambda,T)\sim const\times {T_c^{8/3}\over{\lambda^{1/3}}},
\label{eqn:IIIc}
\end{equation}

If we neglect the contribution of the BLR continuum,
 the continuum flux ratio measured at $\lambda_1$ and
$\lambda_2$ is

\begin{equation}
{F_t(\lambda_1,T)\over{F_t(\lambda_2,T)}}\sim ({\lambda_2\over
\lambda_1})^{1/3}=const.
\end{equation}

There is no evidence that the continuum temperature is proportional
to the BLR temperature, but one can assume that they are in
correlation i.e. that $T_c\sim T_{BLR}$, therefore we will present
the continuum intensity vs. the BLR temperatures.

\begin{figure}
  \begin{center}
    \FigureFile(75mm,75mm){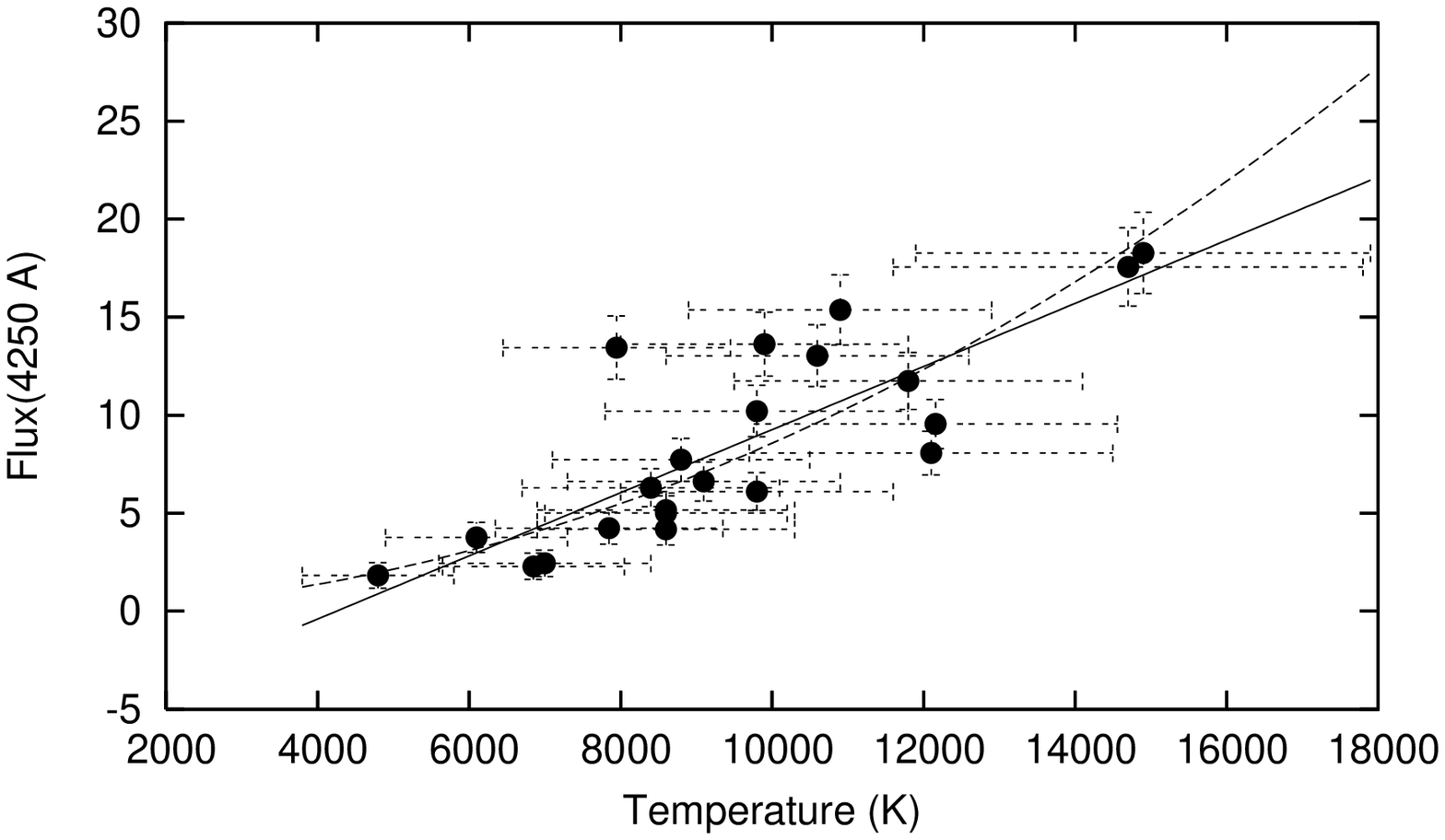}
    \FigureFile(75mm,75mm){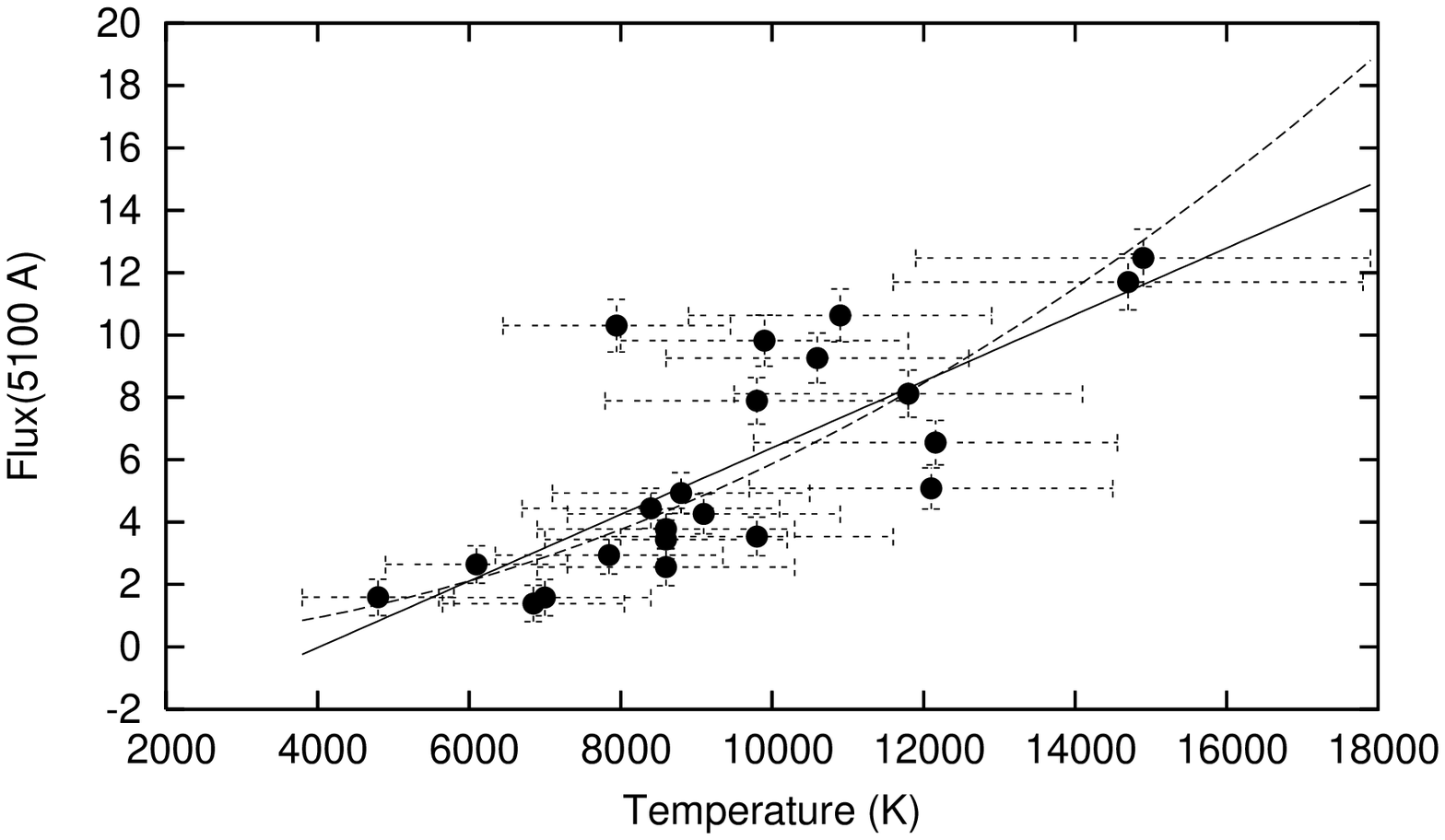}
    \FigureFile(75mm,75mm){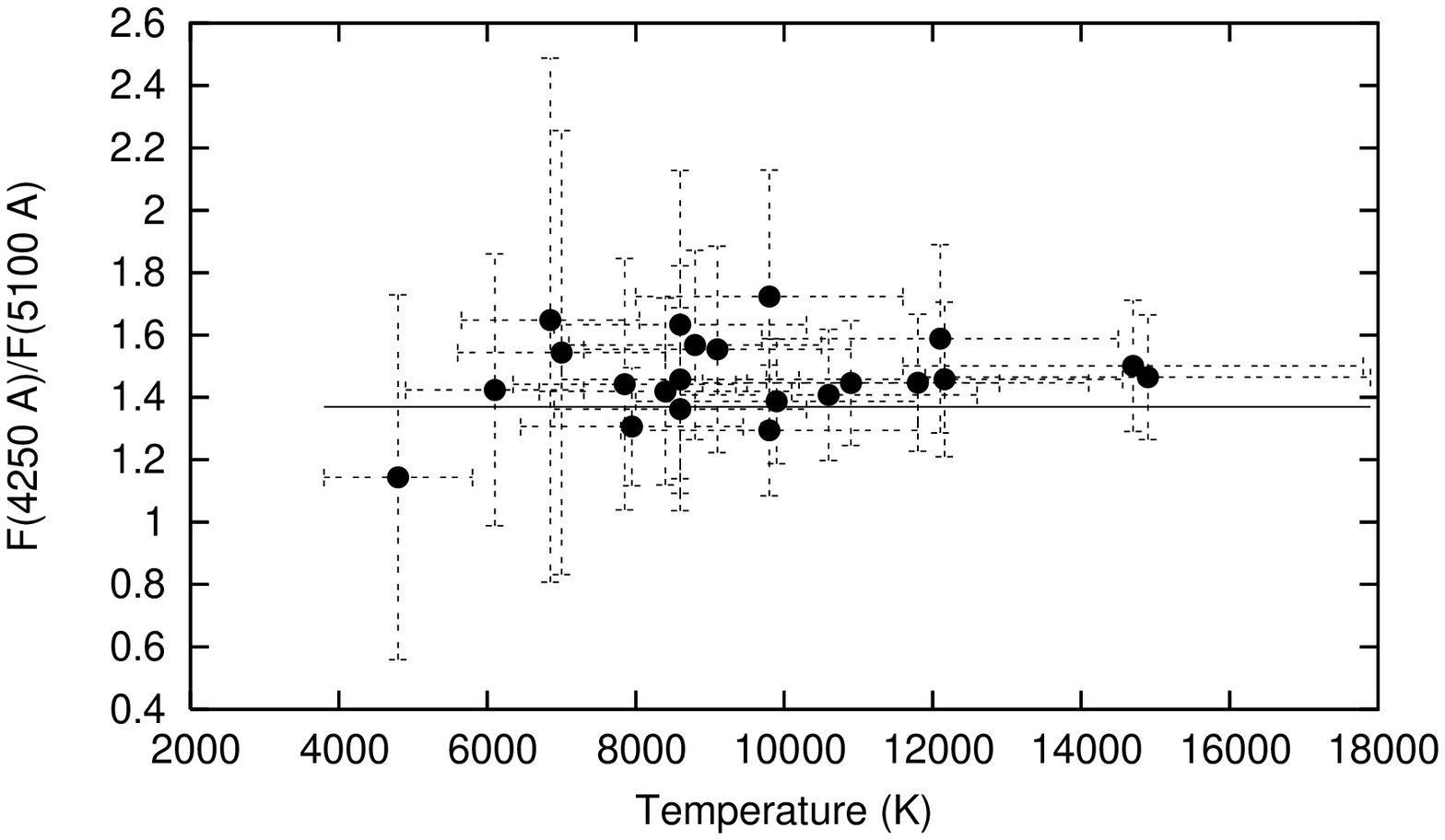}
  \end{center}
\caption{The continuum flux measured at 4250 \AA \ and 5100 \AA\ as
a function of the temperature measured with the BP (first
two panels), and the ratio $I_C(4250)/I_C(5100)$ as a function of
the temperature (bottom).  The measured values are denoted
with full circles. The assumed linear function and the function
given by Eq. (14) are presented with full and dashed line,
respectively (first two panels). In the 3rd panel the full line
represents the best fit with Eq.(15). } \label{fig6}
\end{figure}

 In Figure~\ref{fig6},  we present the AGN-component continuum
flux variation as a function of the measured temperatures (at
$\lambda$4250\AA\ in Figure~\ref{fig6}a and at 5100\AA\ in
Figure~\ref{fig6}b) and the flux ratio variation
(Figure~\ref{fig6}c). As one can see from Figure~\ref{fig6}a,b,
the AGN-component continuum tends to be a linear function of the
temperature. We found a high level of correlation between the
continuum and temperature variability ($r=0.85$). Also, we fitted
the measured data with the continuum function as given by Eq.
(14) and presented it with dashed line in Figure~\ref{fig6}a,b,
finding that Eq. (14) fits well the observed data.

\begin{figure}
  \begin{center}
    \FigureFile(75mm,75mm){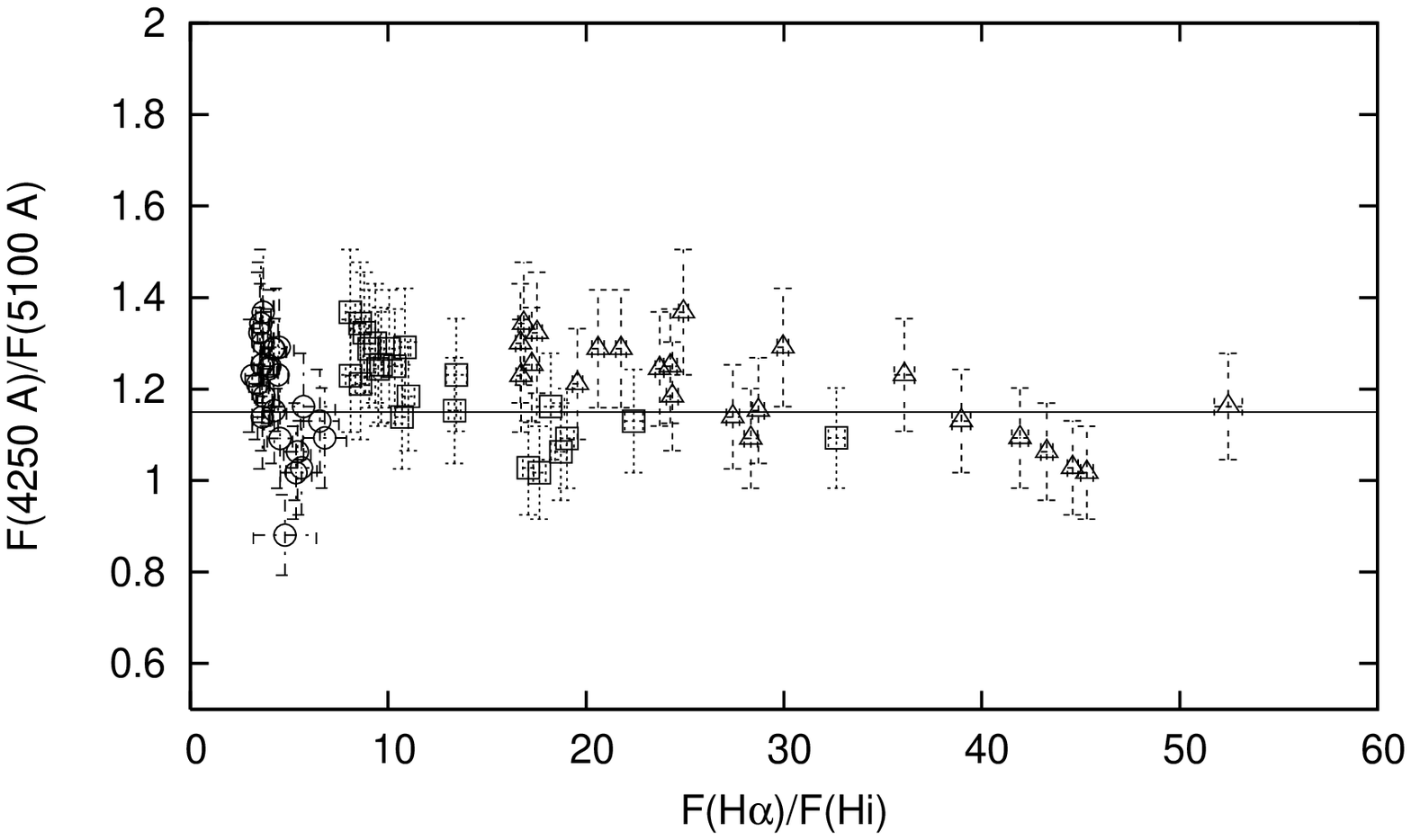}
    \FigureFile(75mm,75mm){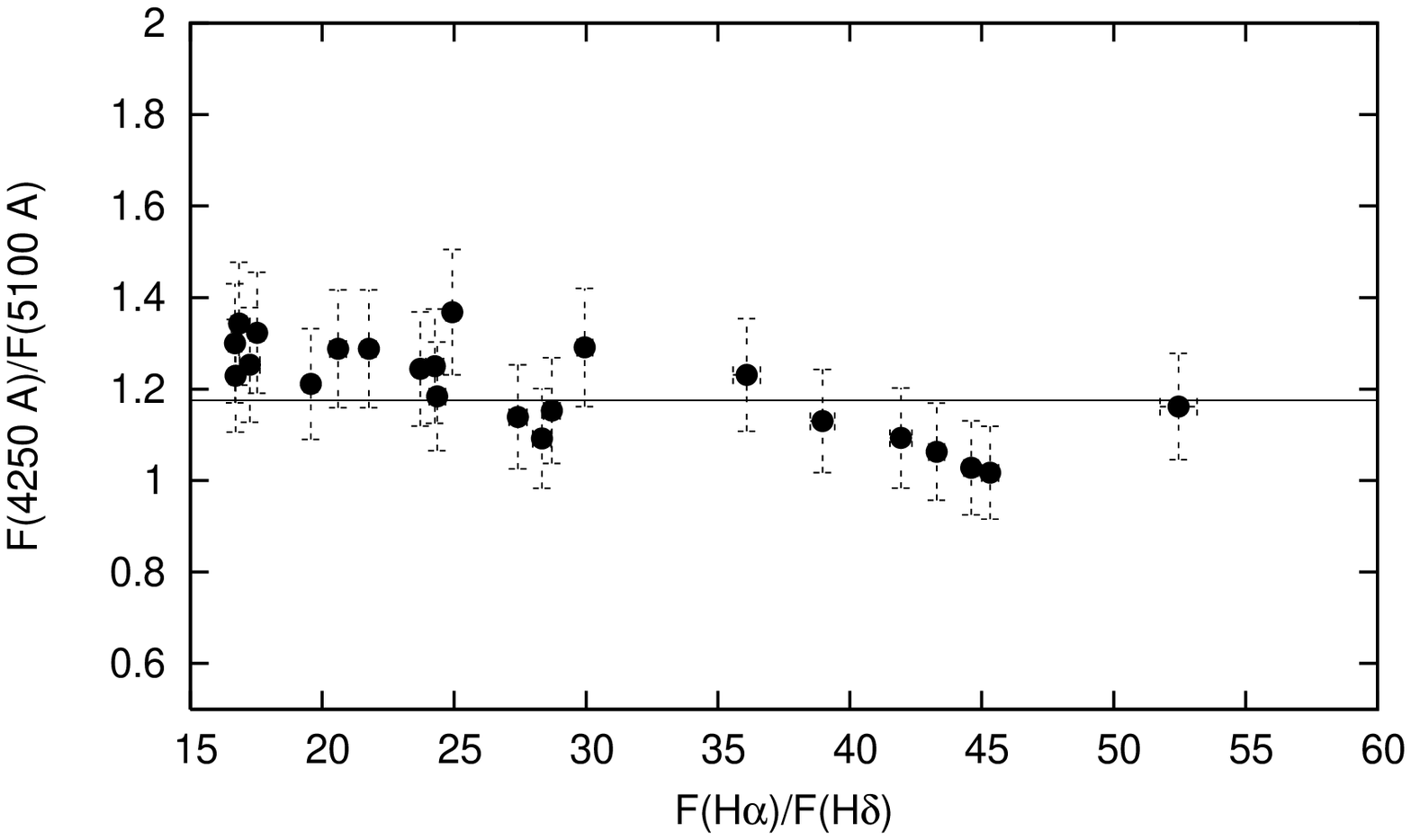}
  \end{center}
\caption{The ratio of measured continuum flux
 as a function of the line flux ratios $H\alpha/H\beta$ (open circles),
$H\alpha/H\gamma$ (open triangles), and $H\alpha/H\delta$ (open
squares). In the bottom panel, the case for $H\alpha/H\delta$ flux
ratio is presented separately. } \label{fig7}
\end{figure}

Furthermore, as one can see from Figure~\ref{fig6}c the AGN-component continuum
intensity ratio at 4250 and 5100 \AA\  follows the expected function
given by Eq. (15), i.e. remains constant as a function of the
temperature. This also indicates the disc emission. But, here we
should note that the temperature measured from the BLR fits very
well the physical changes in the continuum.

This can be expected because of the following reason: whether
illuminated by a variable hard X-ray source at the centre, or by
some other mechanisms, the optical/UV continuum varies with the
ionizing spectrum. Then, one can observe the continuum becoming
"bluer" as it becomes brighter. This is consistent with an increase
in the temperature of the thermal accretion disc as the central
source (X-rays and the ionizing continuum emitting inner disc)
brightens.
The obtained high level of correlation between the AGN-component
continuum and temperature, as well as of the ratio of the
AGN-component continuum at different wavelengths and temperatures,
support a dominant emission by an accretion disc in the BLR. As it
was noted by Ulrich (2000) for the case of NGC 4151 the high
correlation between X-ray emission (originated in an accretion
disc) and temperature of BLR is expected. Here, we showed for the
first time that this correlation exist in the optical part of the
continuum.

On the other hand, the correlation obtained from our measurements
may also be explained by the following: the  emitting gas in the BLR
becomes hotter when the central source brightness and the flux of
incident photons increases. At the same time the sizes of the
Stromgren zones increase in depth, increasing the optical depths in
the Balmer lines. Their emitting efficiencies thus diminish with
increasing incident flux, but the effect on H$\alpha$ is greater
than that on H$\beta$, which is greater than the effect on H$\gamma$
(because H$\alpha$ optical depth remains greater than H$\beta$,
etc., e.g. Ferland et al. 1979). In this case one can expect that
the ratio of Balmer lines fluxes is a function of the continuum
ratio measured in the blue and red part of the spectrum. To check it
we presented the flux ratios of the continuum measured at 4250\AA \
and 5100 \AA\  as a function of $H\alpha/H\beta$, $H\alpha/H\gamma$
and $H\alpha/H\delta$ flux ratios in Figure~\ref{fig7}.
As one can see from Figure~\ref{fig7}, a slight tendency exists for
the line ratios F(H$\alpha$)/F(Hi) to be smaller when the continuum is
bluer, but it is a very tiny effect. It indicates that we have the real
correlation between the temperature and
AGN-component continuum flux, that is presented in Figure~\ref{fig6}.

\section{Conclusions}

By applying the BP method, as it was proposed by Popovi\'c (2003,
2006ab), to the broad Balmer lines of NGC 5548 observed from 1996 to
2004, we found that:

(i)  it seems that the collisional processes play a significant role
and that the distribution of the excited level population ($n>2$)
follows the Saha-Boltzman equation, and this should be taken into
account when modeling  the BLR of NGC 5548;

(ii) the BLR temperature was changed from 5000 K (in 2002) to 15000
K (in 1998). The average temperature is 10000 K, which is a value
expected for the BLR. The maximum of the reached temperature
corresponds to the period where the Balmer lines were the most
intensive;

(iii) we found a correlation ($r=0.85$) between the variation of
optical AGN-component continuum and temperature;
this is the first time that this
correlation is confirmed, and it indicates the presence of an
accretion disc in the BLR of NGC 5548 as it  was earlier suggested
by Shapovalova et al. (2004).


\section*{Acknowledgments}

This work was supported by the Ministry of Science of Republic of
Serbia through the project "Astrophysical Spectroscopy of
Extragalactic Objects" (146002). Also, the work has been financed
by INTAS (grant N96-0328), RFBR (grants N97-02-17625 N00-02-16272,
06-02-16843 and N03-02-17123), state program 'Astron' (Russia) and
CONACYT research grant 39560-F and 54480 (Mexico). We would like
to thank to the anonymous referee for very useful comments.





\end{document}